# ADAPTIVE MULTI-CRITERIA-BASED LOAD BALANCING TECHNIQUE FOR RESOURCE ALLOCATION IN FOG-CLOUD ENVIRONMENTS

*Ahmed A. A. Gad-Elrab[1,2], Almohammady S. Alsharkawy[1], Mahmoud E. Embabi[1], Ahmed Sobhi[1,3], Farouk A. Emara[1]

[1]Department of Mathematics and Computer Science, Faculty of Science, Al-Azhar University, Cairo, Egypt.
[2]Department of Computer Science, Faculty of Computing and Information Technology, King Abdulaziz University, Jeddah, Saudi Arabia.
[3]Faculty of Computers and Artificial Intelligence, AlRyada University for Science and Technology (RST), Egypt.

## *ABSTRACT*

*Recently, to deliver services directly to the network edge, fog computing, an emerging and developing technology, acts as a layer between the cloud and the IoT worlds. The cloud or fog computing nodes could be selected by IoTs applications to meet their resource needs. Due to the scarce resources of fog devices that are available, as well as the need to meet user demands for low latency and quick reaction times, resource allocation in the fog-cloud environment becomes a difficult problem. In this problem, the load balancing between several fog devices is the most important element in achieving resource efficiency and preventing overload on fog devices. In this paper, a new adaptive resource allocation technique for load balancing in a fog-cloud environment is proposed. The proposed technique ranks each fog device using hybrid multi-criteria decision- making approaches Fuzzy Analytic Hierarchy Process (FAHP) and Fuzzy Technique for Order Performance by Similarity to Ideal Solution (FTOPSIS), then selects the most effective fog device based on the resulting ranking set. The simulation results show that the proposed technique outperforms existing techniques in terms of load balancing, response time, resource utilization, and energy consumption. The proposed technique decreases the number of fog nodes by 11%, load balancing variance by 69% and increases resource utilization to 90% which is comparatively higher than the comparable methods.*

## 1. INTRODUCTION

The Internet of Things (IoTs) network has seen a recent surge in the number of linked devices. By 2030, it is anticipated that there will be 500 billion IoTs devices on the planet [1]. Any uniquely recognized gadget with sensing/actuating, processing, and communicating capabilities is considered IoTs device [2, 3]. IoT devices are used to collect data on humans' behalf from sensors about themselves and their surroundings, share data with other devices, process data, and act upon the environment [2]. The global network known as the IoTs is utilized to seamlessly integrate people, processes, and technology through those IoTs devices [1, 4, 5].

There are numerous IoTs applications and services available. Real-time remote patient monitoring in healthcare, smart transport and smart traffic monitoring in smart cities, saving, monitoring and managing power, water or gas use in smart grids, and green computing are examples of IoT applications and services. IoTs encompasses a broad variety of heterogeneous services that have a strong impact on different socioeconomic sectors, each of which has a unique set of needs, such as real-time response, low latency, or high capacity.

Computing and storage resources for IoT devices, including smartphones, are limited [2]. Resources for IoT devices are limited because of the required small size and lightweight for those devices. To overcome this problem, Cloud Computing (CC) [6] is used to provision resources to IoT applications and services. A cloud may consist of a single data centres or a collection of data centers and it can give resources to IoT applications through one or more data centers. A data centre is a collection of clusters of servers with huge computing and storage capacitythat can be accessed through the Internet [7].

Many IoT applications require real-time responses to sensory data. In IoT applications such as catastrophic heart attacks that may occur suddenly to remotely monitored patients, vehicle accidents in a smart transportation app, gas leaks in a smart gas grid, and augmented reality applications, the requirement for real-time response emerges. Indeed, cloud data centresare not ideal for real-time IoT applications due to the Internet's lengthy latency.

To deal with this issue, the Fog Computing architecture (FC) [8] was developed. FC brings the cloud to the edge of the Internet, bringing computing closer to IoT devices. IoT applications become more engaging and latency is considerably reduced by lowering the connection to cloud data centers. A fog in FC might be a single ”fog node” or a network of connected fog nodes. A fog node might be made up of a single server or a cluster of servers, as in [9]. This is known as the Fog-Cloud environment.

In a Fog-Cloud scenario, an IoT application can use resources from both the fog and the cloud. Initially, the IoT application requests resources from the fog. If the available fog resources are insufficient to meet the needs of the IoTs application, the application sends the request to the cloud. Furthermore, cloud data centres can be employed right away to fulfil IoT application requests with tight deadlines and historical data.

The key problem in a Fog-Cloud context is a resource allocation that ensures an efficient load balance among fog nodes. Many load balance techniques and algorithms are used to divide jobs across the available resources of multiple servers in a fog node with a cluster of servers. Some algorithms, such as Round Robin (RR), iterate on the servers, assigning one request to each server per iteration. Other algorithms, such as Weighted Round Robin (WRR) [10], provide a weight to each server based on its total capacity and assign a number of requests to each server based on that weight in each iteration. However, most of these methods do not take the uncertainty of the resource allocation and utilization of a server into consideration.

This paper proposes a load balance technique for resource allocation in a Fog-Cloud environment which takes into account the uncertainty of resource allocation and server utilization. Instead of the number of assigned requests to each server, server total capacity, or number of active connections, the proposed technique defines load balancing in terms of server resource utilization. Furthermore, the proposed technique employs an approach for ranking fog nodes that is based on a hybrid of two multi-criteria decision- making (MCDM) approaches, the FAHPand the FTOPSIS. The FAHP approach is used tocalculate the weights of the criteria, whilst the FTOPSIS method is used to determine the fog nodes ranking. Finally, depending on the nodes' ranking value, the proposed method selects the best node to carry out the current task.

**The main contributions of this paper are as follows:**

- Describing a Fog-Cloud architecture to fit the task requirements of IoT applications.
- Defining load balancing in terms of resource utilization in the Fog-Cloud environment.
- Taking into account multiple objectives such as processing time, response time, and resource utilization of fog devices' in a suitable way to achieve the user and fog devices requirements.
- Formulating the multi-objectives load balancing problem for resource allocation in a Fog-Cloud environment.
- Using FAHP and FTOPSIS for ranking fog nodes in a Fog-Cloud environment. Selecting the optimal node to carry out the current task based on the node ranking value.
- Proposing an adaptive MCDM-based load balancing technique for resource allocation in Fog-Cloud environments.
- An asymptotic performance analysis and extensive evaluation results are carried out to verify the performance of the proposed algorithm.

The rest of the paper is organized as follows: Section 2 introduces a motivation scenario for load balancing in a Fog-Cloud environment. Section 3 includes a detailed survey of the related work. Section 4 describes the proposed Fog-Cloud architecture. Section 5 formulates the load balancing problem in Fog-Cloud environments. Section 6 introduces MCDM Approaches: FAHP and FTOPSIS. Section 7 introduces the proposed load-balancing technique. Section 8 presents a simulation and analysis of the experimental results. Finally, Section 9 concludes the paper.

## 2. MOTIVATION SCENARIO

IoTs have received attention from both industry and academia in recent years, which is advantageous to the daily life of humans. Data gathered from smart sensors is frequently transferred to cloud data centres, and applications are typically processed by processors in the data centres. The IoTs have the potential to benefit medical services applications. In an IoTs-based healthcare system, various types of sensors are employed to measure and monitor various well-being metrics in the human body. These devices can focus on a patient's well-being when they are isolated from others or when the medical facility is out of reach. They can thus provide a real-time answer to the physician, relatives, or patient. These sensors aid in monitoring health parameters such as heart rate, blood pressure, body temperature, respiration rate, pulse, and blood glucose levels. Medical services applications allow the elderly and those with severe medical conditions to live independently and comfortably. IoT advancements contribute to substantial gains in healthcare. Microfluidic biochips and wearable biosensors, for example, can improve clinical diagnoses in a range of settings, from the laboratory to the hospitalShortly, IoTs-enabled gadgets will enable doctors to routinely assess patients with breast, lung, and colorectal malignancies and do point-of-care molecular testing as part of normal care. This will offer physicians with knowledge they need to develop truly data-driven treatment programmes, increasing the chance of a successful recovery.

Cloud infrastructures now in use transport data to cloud servers for further processing before delivering it to devices. CC has been positioned as the primary enabler of IoT applications. This is due to the large capacity and processing power of cloud-based goods, which fully fit the requirements of critical IoT services, such as healthcare or smart transportation services.

CC still has considerable challenges to overcome, some of which are primarily connected to delayed response times, security concerns, and support for global mobility. The main source of these issues is the large distance between the end user device requesting the service and the cloud.

FC, a new architecture, has lately been presented as a solution to these challenges [11]. CC may benefit from fog and edge computing to overcome the limitations of cloud servers.

The main objective of this research focuses on resource management solutions for the Fog-Cloud environment. Given these challenges, good resource management is essential to provide load balancing, reduce latency, and lower the costs of data transit, energy consumption, and storage. The fog layer is utilized to intelligently use resources because it is close to IoT devices, cutting latency and enhancing cloud reliability. The efficacy of resource utilization and load balancing is dependent on selecting the best fog device for the job and meeting user needs. Multiple characteristics must be considered to meet the needs of both users and fog devices.

## 3. Related Work

In this section, we discuss the load-balancing methods, algorithms, and techniques related to different aspects of fog computing.

There are two basic types of load balancing approaches in fog computing: static and dynamic load-balancing techniques [12]. Static load-balancing is based on allocating work based on initial task requirements. These requirements were established at the start of the task. Although this technology is simple to implement and configure, it has considerable disadvantages. They are, for example, increasing the burden of one node during system startup to the maximum load. As a result, task allocation is fixed and cannot be changed while the process is running. Dynamic load balancing, on the other hand, automatically distributes tasks when one of the nodes becomes overloaded. As a result, the destination node is always chosen based on the most current traffic load statistics. As a result, accurate real-time load estimation is critical for active dynamic load balancing. Furthermore, a Virtual Machine (VM) is commonly employed in most computing systems, where it utilises the same pool of resources on the same physical machine [13].

Chandak and Ray [14] presented a survey of load-balancing techniques in fog computing. They also introduced some of the evaluation parameters and simulation tools used for load-balancing methods in the fog system. In addition, Baburao, et al. [15] surveyed some of the techniques of service migration, load balancing, and load optimization in fog computing. Furthermore, Kaur and Aron [16] surveyed load-balancing approaches systematically in a fog environment. Singh et al. [17], [18] conducted a comparison study based on their research regarding different load-balancing approaches, algorithms, and taxonomies. In this context, the authors in [19] introduced an Improved Reptile Search Algorithm (IRSA) to solve the optimization problem which occurs during the time of allocation resources among IoT networks. IRSA employs the methodology of levy flight and cross-over to update the candidate position and enhance the search speed in a single iteration. The proposed method consumes less energy and has low latency during data transmission from User equipment to the base station.

Several methods based on LB are discussed in this survey [20], which overcomes the problem of overloaded data on the network. Latency, bandwidth, deadlines, cost, security, execution time, and execution time are some of the aspects that authors have focused on in LB. Other metrics based on fault tolerance are also addressed, along with their quality parameter table and methodology. In [21], a qualitative and quantitative piece of research was conducted based on an SLR method on load balancing algorithms in fog computing. Applying 1054 studies published recently, between 2013 and 2021, the authors offered the SLR-based method in this literature by using the exploration query. The authors in [17], discuss various kinds of load balancers and their taxonomy by comparing various load balancers, and on the other hand, also discuss their applications. The projected taxonomy might be helpful for researchers and developers to develop their ideas about fog computing.

Zahid et. al. [22] employ a hill-climbing approach to balance loads when computing in the fog. For searching, this method takes advantage of mathematical optimisation. Up until the ideal answer is discovered, this procedure is iterated. In [23], the authors suggest a two-level resource scheduling methodology to reduce the network's response time. The edge, middle, and core levels of the network are divided into the suggested model's new fog computing architecture. The task scheduling is done within the same fog cluster after the resource scheduling model schedules the workloads among different fog clusters. To effectively reduce task delay, a multi-objective optimisation task scheduling technique is also suggested. The suggested improved nondominated sorting genetic algorithm (NSGA-II) aims to increase job execution stability.

In this research work [24], a fuzzy load balancer is devised using different levels of design and tuning of fuzzy controls. This fuzzy logic-based algorithm has been implemented for conducting link analysis as interconnects for managing traffic. Theproposed method [25], controlled every step of the process, from the arrival of the task until its completion. Critical tasks are completed as soon as possible by employing fuzzy logic to assign a priority depending on their predefined priority, deadline, and task size. Fuzzy logic was used to analyse the concept of trustworthiness by Rahman et al [26]. A software-defined component called fog broker was used to calculate reliability. Its primary purpose was to locate fog devices that were readily available and had a solid reputation. Based on feedback, service quality, security considerations, and inputs from the load balancer regarding the current traffic situation, trust and reputation were calculated. Arunkumar and Venkata [27] proposed Feedback based on anoptimized fuzzy scheduling approach (FOFSA). It is divided into two stages: i) The first step is to calculate the fuzzy in the scheduling queue. ii) The second step is to calculatefeedback based on method in the waited queue. The feedback decreases the transmission between IoT devices and the cloud significantly, and it consequently reduces the end-to-end latency automatically.

Secure and sustainable load balancing approach in edge data centre fog computing proposed by Puthal et. al. [28-30]. By processing data streams and user requests in close to real-time, edge data centres, which are situated between cloud data centres and data sources, are well-positioned to reduce latency and network congestion. The author contrasts the suggested solution with a proportional, static, and random task allocation strategy. The proposed algorithm by Chen and Kuehn [31] benchmarked against the Quality of Experience (QoE) for traffic load balancing. The authors anticipated the need for load balancing, especially in cases where dense radio nodes working at the edges of the network [31]. In [32], the authors propose a latency-aware application module management policy to reduce the response time in fog networks.

Shahid et al. [33] presented an energy and delay-efficient fog computing approach with the caching approach. A popularity-based caching method is proposed along with two energy-aware mechanisms. The active fog nodes havebeen chosen based on their energy, power, and number of neighbours. With the filtration approach, the contents are cached on the active node. The efficiency of the cached fog network is increased with the load-balancing approach. This reduces energy consumption and latency, but it doesn't consider the computational cost, so the computational cost is higher. In this context, Kaur et al. [34] and Hussein et al. [35] presented an energy-aware load-balancing approach in fog computing. The first load-balancing approach performed better than existing techniques in terms of time, energy consumption, and cost. But the second has a highpower consumption and a high failure rate. Maswood et al. [36] studied the integration of a fog cloud to reduce the cost of resources and minimize delays in real-time applications and operations. The performance of the proposed model was investigated in terms of links and server utilization, bandwidth cost, and several machines used.The IRSA is utilized in solving the optimization problem in [37]. IRSA is an advancement in the RSA where Levy and crossover methodologies are utilized. The obtained experimental results show that the proposed IRSA attained better performance with an allocation rate.

Shu and Zhu [38] gave a mechanism to offload overheads in 5G networks that had large fog/edge zone-based node deployments. The authors used a term called Weak Load Balancing to demonstrate their algorithm inthe peer-to-peer association of fog nodes. They computed their offloading scheme based ontheminimization of multiple factors that helped to create balance a between network quality and user experience. In [39], the authors presented the P2PFaaS framework, a software suite which enables the testing and benchmarking of scheduling and load-balancing algorithms among sets of real nodes. [40] this paper explores recent articles to determine the possible research gaps and opportunities to implement an efficient solution for load balancing in fog environments after analyzing and assessing the existing solutions. In this study [41], the authors proposed a secure and energy-aware fog computing architecture,and implemented a load-balancing technique to improve the complete utilization of resources with an SDN-enabled fog environment.

The load may become imbalanced because some fog nodes don’t receive the adequate amount of resources as a result of improper resource scheduling. Scheduling resources for idle fog nodes will also cause a load imbalance and a loss of power. The conventional systems have handled load balancing in a wide variety of methods. However, the changing nature of tasks and their urgency in fields like medicine have not been taken into account. Load balancing in a fog computing system must manage both the task and the entire process. It requests permission to access the fog nodes to complete the task. Conventional solutions have not been able to handle the scenario because load balancing considers task execution and resource allocation at their operating stage. Here isa list of some advantages and disadvantages of previous load balancing in IoTs Fog computing environments, high response time, high computational, high failure rate, high power consumption, high task execution time, high communication cost, consumption of a lot of energy, priority of task, current fog node utilization, and load balancing and task scheduling need a long processing time. A few researchers have studied load balancing via managing resource distribution. By taking into account the tasks’ requirements, fog node utilization and available resources, the proposed load balancing method has concentrated on offering an appropriate resource allocation utilizing multiple objectives to distribute the tasks to the appropriate fog node and achieving load balancing between the current fog nodes.

## 4. THE FOG-CLOUD ARCHITECTURE

Multiple channels are used in the architecture of FC to generate data from IoT devices. In a cloud environment, this data collection can be analyzed using analytical tools. Transferring large amounts of data to the cloud has an impact on factors like transfer rate, bandwidth, and latency, among others. Fog computing enables data to be processed near to where it is being generated rather than sending allofit to the cloud for processing. By processing data close to the client, latency, bandwidth, and even data transport costs can be reduced. The data obtained from the fog layer is kept for a long time in the cloud layer where it is stored.

The Fog-Cloud architecture consists of three layers: an IoTs layer, a fog layer, and a cloud layer. The architecture allows for wired or wireless connectivity between the IoTs, fog, and cloud layers. The following is an explanation of these three layers:

**IoTs layer:** Wearable gadgets, smart home sensors, actuators, and healthcare equipment are all included in this layer via a network. These gadgets and sensors are able tocan sense, gather, and communicate data in real-time through networks. This layer also incorporates complex algorithms, cloud interfaces, and communication interfaces in addition to these devices. These components work together to create a communication network, and the data they produce is sent to the cloud via the fog layer. The IoT framework’s backbone is created by the sensing network. The main purpose of the IoT architecture is to collect data about an object (such as a patient’s health) and distribute it via wireless channels to higher levels. The implementation of this IoTs framework makes use of particular protocols suchas WiFi, Bluetooth, ZigBee, etc.

**Fog layer:** This layer is closer to the IoTs layer and serves as an intermediary layer between the IoTs and cloud layers. At the network's edge, the fog layer delivers low-latency computing services. Local resources can be effectively used by fog-enabled network architecture and services to support delay-sensitive IoT applications in regional environments. It decreases traffic transmissions and centralized computer requirements, improving overall network throughput performance. The fog layer consists of processing devices, gateways, and networked devices distributed between network edges and clouds. Each fog device has a limited computational, networking, and storage capability. The fog computing layer receives data from various IoT sensors and devices. Some tasks with the highest time urgency and lowest computation density may be processed in the fog layer rather than in cloud layer. In addition, the fog layer is utilized for real-time processing and analysis of IoT data, and it is linked to the cloud layer for storage and analysis of results.

**Cloud layer**: This layer has many physical data centre nodes that communicate with each other via Wide Area Networks (WANs). Each cloud data centre has its own set of hardware configurations (memory, CPU, network bandwidth, capacity, and storage) that are used to meet customer requests for necessary resources. The cloud layer is in responsible for storing, processing, and executing tasks that the fog layer cannot process and execute. Fig.1 shows a high-level overview of the Fog-Cloud architecture

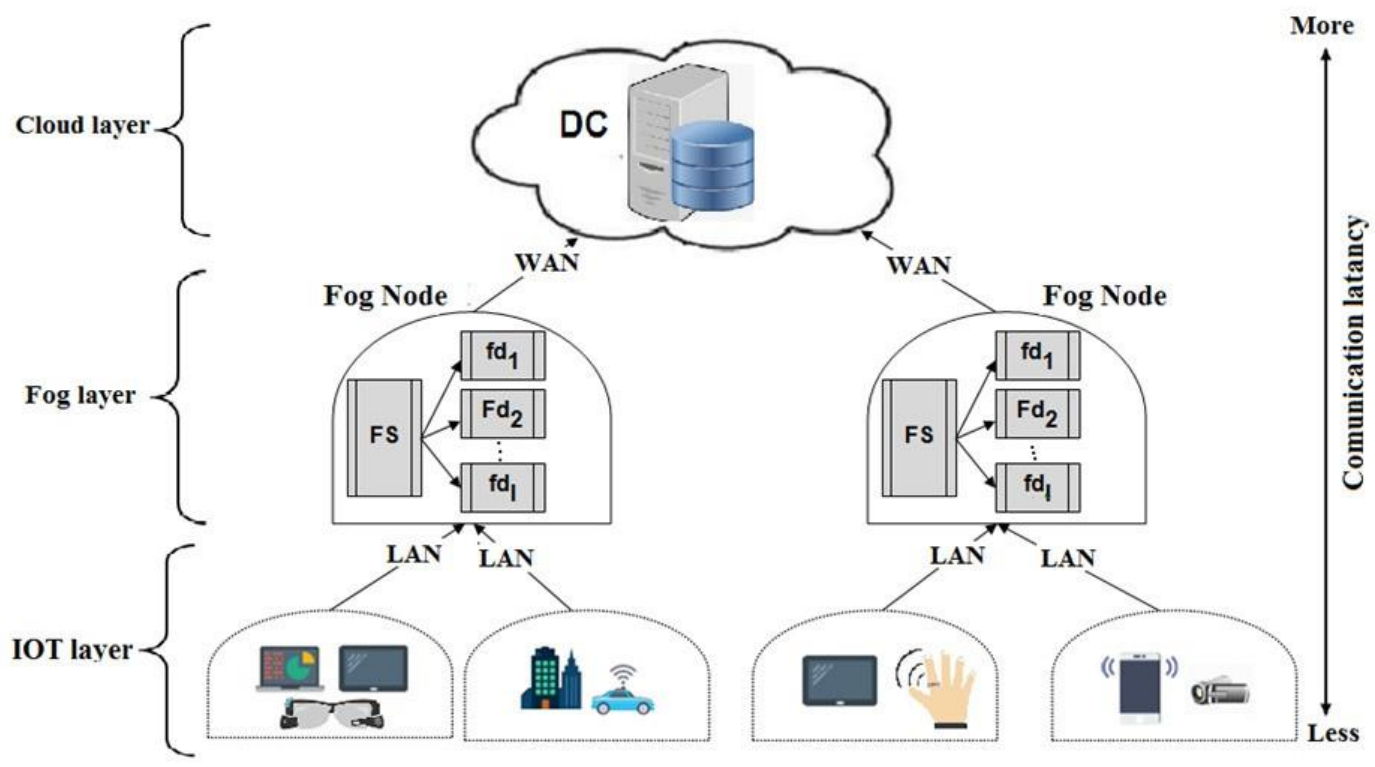

Figure 1.The Fog-Cloud Architecture

Based on the described architecture infigure.1, the workflow for executing tasks based on Fog-Cloud architecture is illustrated as follows.

1. Initially, when an IoT device or a user sends its task, it will be forwarded to the fog layer for processing.
2. The tasks with low computing resources are being processed at the fog layer.
3. The tasks with greater computing resources will be forwarded to the cloud layer for processing.
4. The fog server will start to perform tasks based on the current status of each fog device and the requirements of the task.
5. Each fog device receives a list of tasks to perform, when a task is completed, the fog device will send the task outcomes to the fog server and the fog server forwards the output to the corresponding IoT device. Also, send the processed data and other application-related data to the cloud data center for long-term and future use. Figure.2 illustrates the general procedures for executing tasks using the proposed Fog-Cloud architecture.

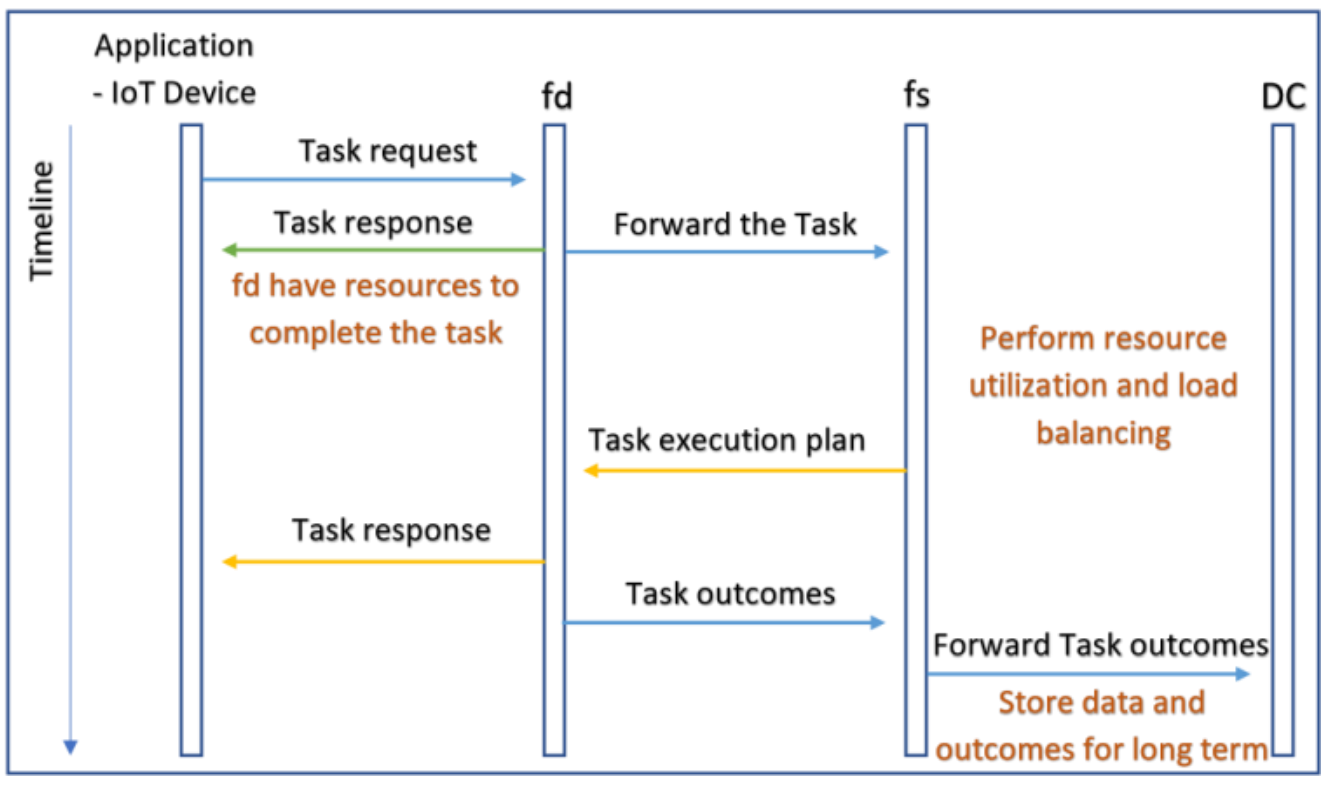

Figure 2.The general steps for executing tasks by using the proposed Fog-Cloud architecture

## 5. Load Balancing Problem Formulation In Fog-Cloud Environment

In this section, the problem of resource utilization and load balancing in a fog-cloud environment will be formulated. Firstly, models, assumptions, and definitions will be provided to help in clarifying this problem.

### 5.1. Models, Assumptions and Definitions

For resource utilization and load balancing in the Fog-Cloud context, the following components: a set of End Devices, Fog nodes, and Cloud servers will be evaluated.Let $O$ be a set of IoTs devices, where $O = \{o_1, o_2, ..., o_n\}$ where $n$ is the number of IoT devices. The IoTs device collects data and sends it to the fog node for processing as a task based on the application request, which needed to be performed as soon as possible. Let $T$ indicate the set of tasks which will performed by the fog nodes, $T = \{t_1, t_2, ..., t_k\}$, where $k$ is the number of tasks. Each task $t_i$characterized by response time $t^{RT}{}_i$, required processing $t^P{}_i$, required cache memory $t^C{}_i$, required memory $t^M{}_i$, required bandwidth $t^B{}_i$and required storage $t^S{}_i$. The proposed model contains a set of Fog Nodes $FN$, where $FN = \{fn_1, fn_2, ..., fn_g\}$, which is responsible for receiving all tasks from the near IoTs devices, where $g$ is the number of fog nodes. Each fog node has a number of Fog Devices FDs, these FDs process the tasks $T$, where $FD = \{fd_1, fd_2, ..., fd_l\}$. Each $fd_j$is characterized by available processing, available cache memory, available memory, available bandwidth, available storage, and available Energy and these characteristics represented as a vector

$$\beta_j = \langle fd_j^P, fd_j^C, fd_j^M, fd_j^B, fd_j^S, fd_j^E \rangle$$

The proposed technique also contains a fog server $FS$ which initially is a fog device belonging to a Fog Node and has a higher configuration compared with other FDs. Processing capabilities (CPU) $fs^P$, memory $fs^M$, storage $fs^S$and network capacity $fs^N$should be higher in this device compared with the FDs in the Fog Node. Each $fs$ manages a cluster of FDs and isresponsible for orchestrating the work between all the FDs. The FS is responsible for organising the resource utilization and the selection of FDsandalso maintaining load balancing among different FDs in the same cluster. Finally, the proposed technique considers the cloud server $CS$, which will keep all the stored processed data and other application-related data for the long term. Each cloud server $CS_i$manages a set of data centre$DC = \{dc_1, dc_2, ..., dc_q\}$, where $q$ represents the number of data centres.

In this paper, the Fog devices Load Balancing is a mechanism that enables jobs to move from one computer to another within the distributed system. This creates faster job service e.g., minimizes job response time and enhances resource utilization. Load balancing among fog devices of a distributed system highly improves system performance and increases resource utilization. Load balancing is the process of roughly equalizing the work-load among all fog devices of the Fog-Cloud architecture. The $X_{ij}$is a binary variable to judge whether the task $t_i$is assigned to the fog device $fd_j$, which is calculated as follows:

$$X_{ij} = \begin{cases} 1, \text{if } t_i \text{ is assigned to the fog device } \mathrm{fd}_j \\ 0, \quad otherwise \end{cases} \quad (1)$$

The main goals of this paper are reducing task response time while taking into account fog device resource utilization and load balancing among all fog devices in the cluster and reducing the cost of task execution. These objectives will be met by employing an efficient method for allocating each task to the best FD that meets the requirements of applications and FNs. These requirements such as maximizing resource utilization and achieving load balancing between different fog devices in the fog node, as well as minimizing the task execution consumed energy to reduce the environmental impact of device energy use, we will detail the objective and constants of this research.

## 5.2. Task Processing Time Cost

To compute the task processing time $PT^t$, we need to obtain two values for each task, the first value is the size of the data which will be processed $t^d$ [42], andthe second value is the number of the task's instruction $t^l$[6]. The two values will be used to compute the data processing time $PT^d$ and the task's instructions processing time $PT^l$as follows:

$$PT_{ij}^{d} = \frac{t_i^d}{v_j} \quad (2)$$

$$PT_{ij}^{l} = \frac{t_i^l}{\rho_j} \quad (3)$$

Where $t^d_i$the amount of request's data to be processed, $v_j$the speed of the computing cores at $fd_j$, $t^l_i$is length of task (given in MI), $\rho_j$is the processing speed (given in MIPS) of $fd_j$. The total tasks processing time $PT_i^t$for task $t_i$is the sum of data processing time Eq.2 and instruction processing time Eq.3.

$$PT_{ij}^{t} = PT_{ij}^{d} + PT_{ij}^{l} \quad (4)$$

## 5.3. Task Response Time Cost

The task's Response Time $t^{RT}_{ij}$is the amount of time required for responding to a user's task. In other words, the operating speed of the system is measured by its response time. It can be computed as follows:

$$t_{ij}^{RT} = PT_{ij}^{t} + t_i^{dep} \quad (5)$$

Where $PT_{ij}^{t}$ is the total tasks processing time for task $t_i$is the sum of data processing time and instruction processing time Eq.4. The task's deployment time $t^{dep}_i$takes into account the time elapsed before the proper fog device of each task on the fog node. This time include, the communication time (sending the application's instructions and the data to be processed), the time eclipsed by fog server to distribute the tasks and load balancing phases.

### 5.4. Resource Utilization Cost

The efficient resource utilization of fog nodes can reduce the required bandwidth betweenfog and cloud computing, decrease the fog node's energy consumption, decrease the latency, thereby meeting the task deadline. The resource utilization of fog devices ($RU_j$) represents the number of tasks handled by this fog device in the available time. The value of $RU_j$can help the fog resource manager to obtain completed information about the resources and their utilization. Also, it allows the manager to make most of the fog resources available to execute tasks within their deadlines and decrease the tasks that are sent to cloud computing. It can be computed as follows:
The resource utilization ratio for running task $i$on $fd_j$is formulated as follows:

$$\eta_{ji} = \frac{PT_{ij}^{t}}{fd_j^{time}},\ \forall\ j,\ s.t\ t_i \in E \qquad (6)$$

Where $fd_j^{time}$ is the available time of $fd_j$which includes the instruction and data processing time, and $E$ is a set of $y$ tasks which selected to be run on $fd_j$. The selected $fd_j$to perform $t_i$has the highest weight among all fog devices $max(fd_j^{w})$. The total resource utilization ratio for $fd_j$is formulated as follows:

$$RU_j = \frac{\sum_{i=1}^{y} \eta_{ji}}{y} \qquad (7)$$

### 5.5. Energy Consumption Cost

In this research, we use $E_c(fd_j)$ to denote the power consumption for executing task $t_i$at a fog device as shown in Eq.8. $f_j$ is a superlinear function denoting the CPU energy consumption of device $fd_j$. $\beta$ is a pre-configured model parameter depending on the chip architecture [43–45].

$$Ec_i(fd_j) = \beta \cdot f_j^3 \qquad (8)$$

### 5.6. Work Load Cost

Let $fd_j^{ld}$denote to load factor of $fd_j$, where load factor is a percentage represent the overall $fd$resources ($fd_j^P, fd_j^C, fd_j^M, fd_j^B, and fd_j^S$) denoted as $fd_j^{\alpha O}$ and the allocated resources ($t_i^P, t_i^C, t_i^M, t_i^B and\ t_i^S$) of the assigned tasks to a $fd_j$denoted as $fd_{ji}^{\alpha A}$. Eq.9, represents the allocation ratio of each resource for $fd_j$, Eq.10, represents the allocation of all $fd_j$resources, and Eq.11, represents the total load of $fd_j$.

$$fd_{ji}^{\alpha P} = \frac{t_i^P}{fd_j^P}, fd_{ji}^{\alpha C} = \frac{t_i^C}{fd_j^C}, fd_{ji}^{\alpha M} = \frac{t_i^M}{fd_j^M}, fd_{ji}^{\alpha B} = \frac{t_i^B}{fd_j^B}, fd_{ji}^{\alpha S} = \frac{t_i^S}{fd_j^S} \qquad (9)$$

$$\boldsymbol{fd_{ji}^{\alpha A} = fd_{ji}^{\alpha P} + fd_{ji}^{\alpha C} + fd_{ji}^{\alpha M} + fd_{ji}^{\alpha B} + fd_{ji}^{\alpha S}} \qquad (\mathbf{10})$$

$$fd_j^{ld} = \sum_{i=1}^{y} fd_{ji}^{\alpha A} \cdot 100. \qquad (11)$$

### 5.7. Problem Formulation

In this paper, the problem of fog device load balancing is defined as a selection problem based on multi-criteria. Here, the cost of task execution is considered as the main objective, which is the sum of the task response time and the consumed energy from executing the task on an$fd_j$. The task execution cost will be computed as follows:

$$TC_{ij} = \left[\frac{t_{ij}^{RT}}{max(t_{ij}^{RT})} + \frac{Ec_i(fd_j)}{max\left(Ec_i(fd_j)\right)}\right] \quad (12)$$

To normalize the values of the task's response time and fog device energy consumption, both of themaredivided by maximum response time and maximum energy consumption, respectively.

With these observations, the problem of minimizing the task execution cost can be formulated as follows:

$$minimize \sum_{i=1}^{k}\sum_{j=1}^{l} TC_{ij}\, X_{ij} \quad (13)$$

Subject to

$$\sum_{j=1}^{l} X_{ij} = 1,\ \forall\ t_i \in T, \quad (14)$$

$$t_{ij}^{RT} \le t_i^{dt}, \quad (15)$$

$$RU_j \neq 0, \ \ , \quad (16)$$

$$fd_r^{ld} \approx fd_j^{ld},\ \forall\ fd_r\ and\ fd_j \in FD. \quad (17)$$

Constraint 14 means that each task $t_i \in T$ will be assigned to one and only one fog device. In Constraint 15, $t^{dt}_i$is the deadline time of the task $t_i$, this condition means that the task response time must not exceed the task deadline time. Constraint 16 means that each fog device must execute a task (tasks). Constraint 17, represents the main objective of this research which is achieving the load balancing among all fog devices. According to the definition of Fog devices Load Balancing, to achieve the load balancing between all fog devices, the fog server must distribute all the received tasks to the available fog devices, such that the workload of all fog devices is roughly equalized.

# 6. MULTI-CRITERIA DECISION MAKING APPROACHES

This section introduces the Fuzzy TOPSIS (Technique for Order of Preference by Similarity to Ideal Solution) and the Fuzzy Analytic Hierarchy Process (FAHP). The most intuitive and simple methods to handle multiple criteria decision-making (MCDM) problems can be regarded as FAHP and TOPSIS.

## 6.1. Fuzzy Sets and Fuzzy Numbers

The Fuzzy Set Theory (FST) was developed by Zadeh in 1965 to address the ambiguity and uncertainty of data. FST has a lot of benefits, including the ability to represent uncertain data. FST also enables the application of mathematical operations and programming to the fuzzy domain. A class of objects known as a fuzzy set (FS) has a continuum of membership grades. A membership function that awards each object a membership grade that falls ”between” zero and one defines such a set.

*fuzzy Set*: A fuzzy set $\tilde{A}$. In a universe of discourse X is characterized by a membership function $\tilde{\mu}_A\ (x)$ which associates with each element x in X a real number in the interval [0, 1]. The function value $\tilde{\mu}_A\ (x)$ is termed the grade of membership of x in $\tilde{A}$. L.A. Zadeh [46].

*Triangular Fuzzy Number*: A triangular fuzzy number $\tilde{A}$can be defined by a triplet (L, M, U) shown in Figure.3. The membership function $\tilde{\mu}_A\ (x)$ is defined in [47] as

$$A = \begin{cases} 0 & x < L \\ \dfrac{x-L}{M-L} & L \leq x \leq M \\ \dfrac{x-U}{M-U} & M \leq x \leq U \\ 1 & x > U \end{cases} \quad (18)$$

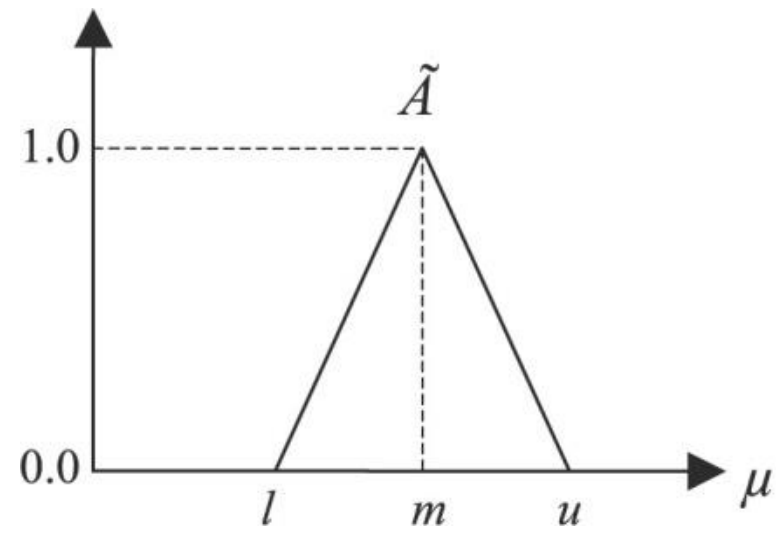

Figure 3.Triangular Fuzzy Number (L, M, U)

A Symbol ”∼” will be placed above A if the A shows a FST. A Triangular Fuzzy Number (TFN) A, $\tilde{A}$TFN represented with three points as follows: (L, M, U). L stand for the lower bound of the fuzzy number and U stand for the upper bound. This representation is interpreted as membership functions and holds the following conditions.

- L to M is increasing function
- M to U is a decreasing function
- $L \leq M \leq U$.

The idea of fuzzy sets was initially brought up to address issues with subjective uncertainty. The use of linguistic variables conveyed by verbal words or phrases in a natural or artificial language—to represent the issue or the event leads to subjective uncertainty. To assess the fulfilment of the performance value for each criterion, linguistic variables are also used. Since the corresponding membership function and the fuzzy interval can be used to determine the linguistic variables. Linguistic variables were proposed in [48], For example, linguistic variables with triangular fuzzy numbers may take on effect values such as very high (very good), high (good), fair, low (bad), and very low (very bad). So, we can naturally manipulate the fuzzy numbers to deal with the FMADM problems. The membership function of linguistic variables represented in triangular fuzzy numbersis showed in Figure.4.

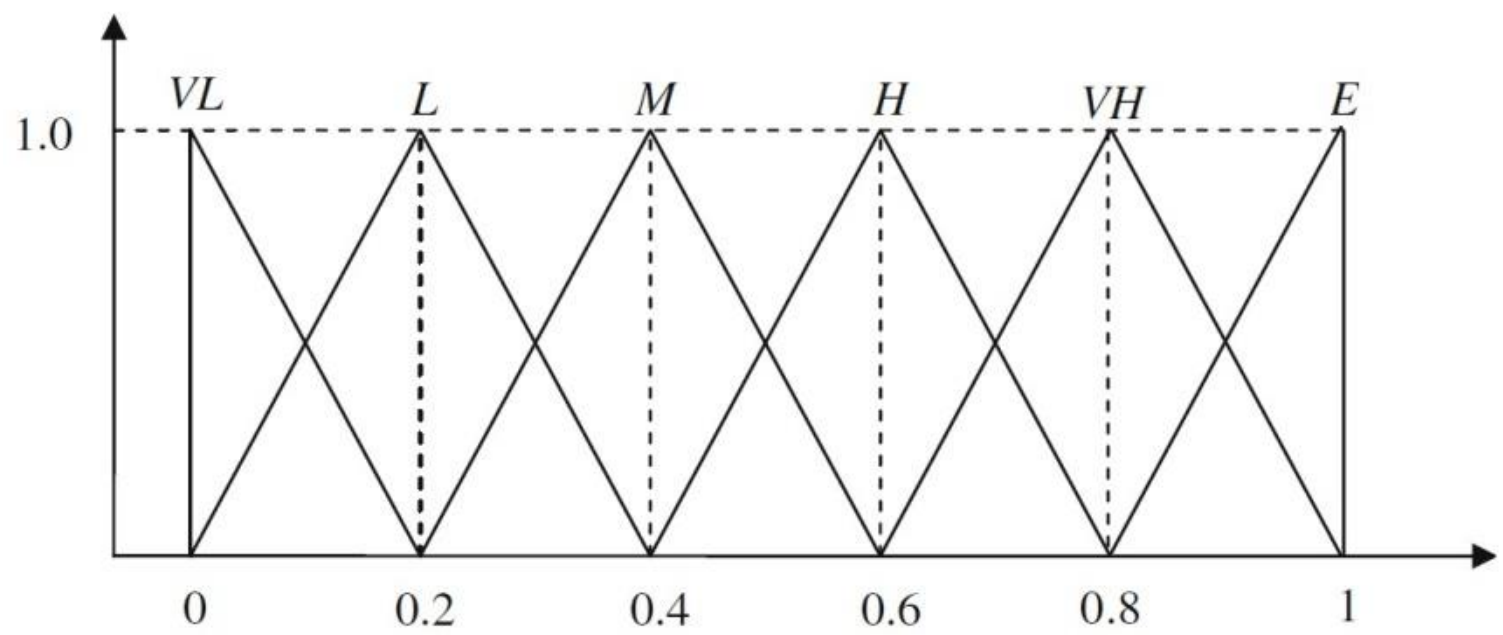

Figure 4.Triangular fuzzy numbers of linguistic variables

### 6.2. Fuzzy Analytic Hierarchy Process (Fahp)

Bernoulli (1738) proposed the concept of utility function to reflect humanpursuit, such as maximum satisfaction, and von Neumann and Morgenstern (1947) presented the theory of game and economic behaviormodel, which expanded the studies on human economic behaviour for multiple criteria decision-making (MCDM) problems [48], an increasing amount of literature has been engaged in this field.

### 6.3. Fuzzy Technique For Order Of Preference By Similarity To Ideal Solution (Ftopsis)

Hwang and Yoon (1981) proposed the Technique for Order Preferences by Similarity to an Ideal Solution (TOPSIS). The fundamental notion was derived from the concept of the compromise solution, which is to select the best alternative with the least Euclidean distance from the positive ideal solution (optimal solution) and the greatest Euclidean distance from the negative ideal solution. Positive-ideal solutions (PIS) maximize the benefit criteria while minimizing the cost criteria, whereas negative-ideal solutions (NIS) maximize the cost criteria while minimizing the benefit criteria. Then, choose the best sorting result as the best alternative. As a result of this technique, we can evaluate mobile nodes based on their context.

The use of numerical values (Crisp values) in the rating of alternatives may have limitations to deal with uncertainties and ambiguity. So, extensions of *TOPSIS* were developed to solve problems of decision-making with uncertain data resulting in F*TOPSIS*. In practical applications, the triangular shape of the membership function is often used to represent fuzzy numbers. Fuzzy models using triangular fuzzy numbers proved to be very effective for solving decision-making problems where the available information is imprecise.
Given a set of alternatives, $A = \{A_k \mid k = 1, ..., n\}$, and a set of criteria, $C = \{C_j \mid j = 1, ..., m\}$, where $X = \{X_{kj} \mid k = 1, ..., n, j = 1, ..., m\}$ denotes the set of performance ratings and $w = \{w_j \mid j = 1, ..., m\}$ is the set of weights, the obtained information from FAHP technique.

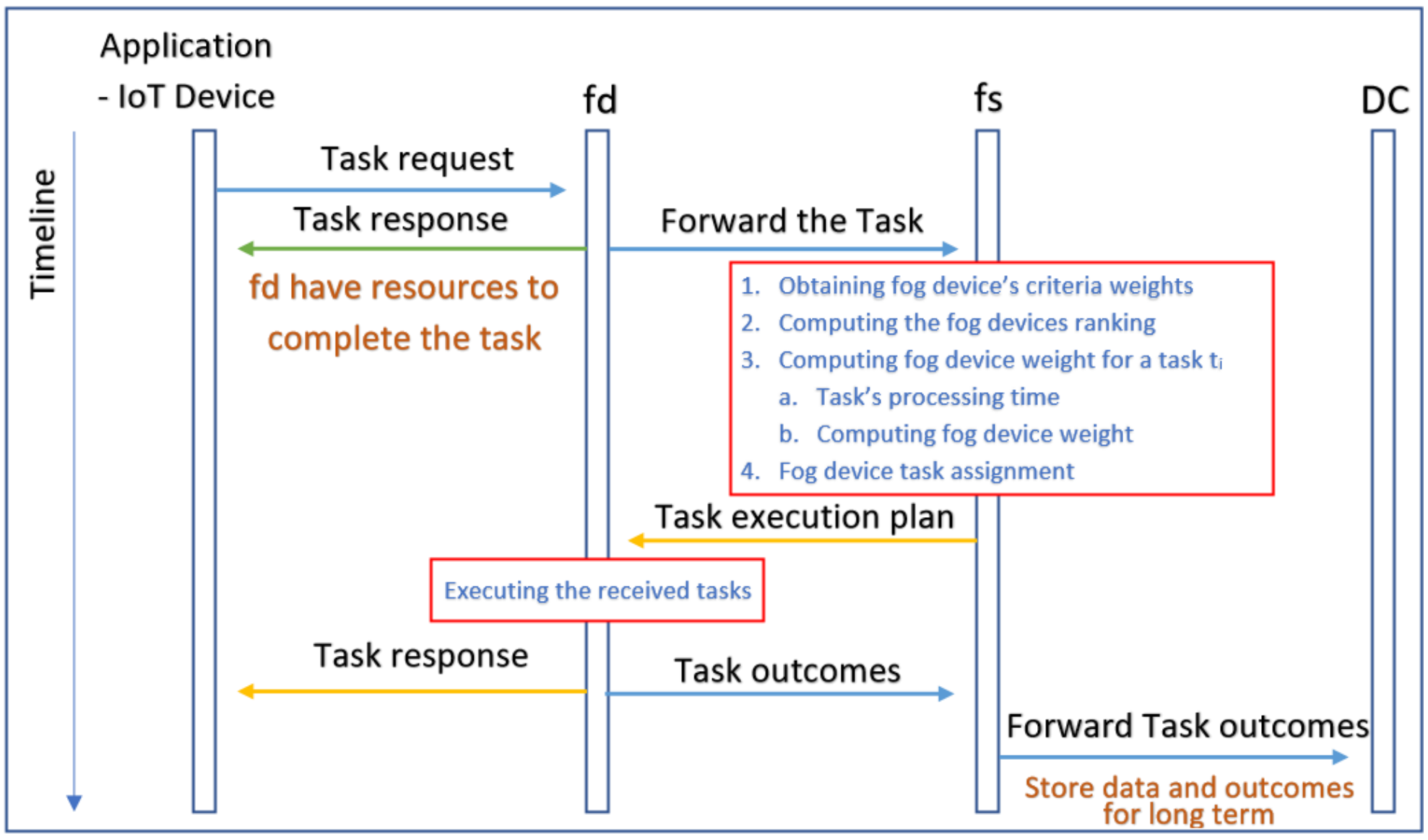

Figure5. The Work flow of the proposed AMCLBT

## 7. THE PROPOSED LOAD BALANCING TECHNIQUE FOR RESOURCE ALLOCATION IN FOG-CLOUD ENVIRONMENT

To address the load balancing issue in the fog-cloud environment, a technique known as the Adaptive Multi-Criteria-Based Load Balancing technique (AMCLBT) was developed. The basic idea of AMCLBT is based on ranking each fog device based on available resources and then determining the weight of each fog device for the current task. Then select the highest fog device weight to carry out the current task. Figure.5, describes the work-flow of the proposed load balancing technique. The proposed AMCLBT consists of four phases as follows:

Table 1.The obtained *fd*criterion weights by FAHP

| **Criterion** | **Weight** |
|---|---|
| $C_1$ (Available Processing) | 0.2821 |
| $C_2$ (Available Cache memory) | 0.1095 |
| $C_3$ (Available Memory) | 0.2153 |
| $C4$ (Available Bandwidth) | 0.1722 |
| $C5$ (Available Storage) | 0.0689 |
| $C6$ (Available Energy consumption) | 0.1521 |

**Phase 1: Calculating Criteria Weights**

Evaluating an*fd*based on its available resources is a multi-criteria decision-making problem. In this phase, the FAHP will be used to obtain the weight of each *fd*criterion. Based on the mentioned *fd*criteria, a matrix to calculate a set of pairwise comparisons will be constructed. The comparison is made through a scale to show "how many times more important or dominant one criterion is over another criterion" [48]. The matrix in Eq.19 represents the pairwise comparison of the selected *fd*criteria, then the equations of FAHP will be used to obtain the weight of each criterion see Table 1.

$$\tilde{A} = \begin{bmatrix} 1 & 3 & 2 & 2 & 1 & 3 \\ 0.33 & 1 & 3 & 1 & 3 & 2 \\ 0.5 & 0.33 & 1 & 2 & 3 & 2 \\ 0.5 & 1 & 0.5 & 1 & 2 & 3 \\ 1 & 0.33 & 0.33 & 0.5 & 1 & 2 \\ 0.33 & 0.5 & 0.5 & 0.33 & 0.5 & 1 \end{bmatrix} \qquad (19)$$

**Phase 2: Ranking Fog Device**

In this phase, the *fs* uses the computed fog devices criteria weights to rank all fog devices in the fog node. The FTOPSIS method is used to rank each fog device according to the specified criteria. The ranking method will use six criteria as mentioned previously, all these criteria are benefits criteria except the energy consumption which is a cost criterion. The AMCLBT uses the equation of relative closeness to the ideal solution to compute the rank $fd_j^r$ for each fog device [48].

**Phase 3: Determining Fog Device Weight**

In this phase, the *fs* will compute the weight for each $fd_j$. This weight is based on two parameters, the first one is the rank of a fog device$fd_j^r$, the second is $PT_{ij}^t$ the total task's processing time on each $fd_j$. The computation of $fd_j^w$based on the fog device's rank and the task's processing time as follows.

$$fd_j^w = fd_j^r * w_q + \frac{1}{PT_{ij}^t} * w_e \qquad (20)$$

where $fd_j^r$ is the rank of $fd_j$, $w_q$ and $w_e$ are the weight of fog device rank and the weight of task processing time respectively, such that $w_q + w_e = 1$.

**Phase 4: Selecting Fog Device**

In this phase, the *fs* will assign the current task $t_i$ to the highest weight $fd_j$. In addition, the *fs* will update the FDs resource utilization table to keep up with the current situation of resources.

The previous phases will be repeated for all the received tasks. From the previous phases, the proposed technique can achieve the load balance between all fog devices in the fog node. The load balancing achieved by assigning each task to the highest-weightfog device which will maximize resource utilization and decrease the response time of a task. Algorithm 1 illustrates the steps of AMCLBT.

**Algorithm1** AMCLBT

**Input:** list of *k* Tasks, list of *l* fog devices, fog devices characteristic cs

**Output:** vector of each task and the selected *fd*

Calculate the criteria weights using FAHP  
**for** $i = 0$ to $k$ **do**  
    Rank FDs using FTOPSIS  
    **for** $j = 0$ to $l$ **do**  
        Calculate Task's processing time $PT_i^t$ for $fd_j$. Eq.4  
        Calculate $fd_j$ weight using Eq.20  
    **end for**  
    assign the task $t_i$ to the highest weighted fog device $fd_j$  
    Add task $t_i$ and $fd_j$ to the tasks' execution vector  
    Update $fd_j$ characteristics  
**end for**

## 8. SIMULATION AND RESULTS EVALUATION

This section presents the simulation results and analysis of the proposed load-balancing technique for IoT devices in a fog environment.

Here areseveral experiments to study the performance of the proposed AMCLBT. The standalone Java code on a device with specifications (core i7 7th and ram 8 GB) is used to implement AMCLBT and analyze the performance of DRAM [49], QLBA [50], and the proposed AMCLBT technique. The performance is measured in terms of average number of fog nodes, average load balancing variance, average resource utilization, and average turnaround time for heterogeneous and homogeneous tasks. In these experiments, the AMCLBT is implemented by using different numbers of nodes ranging from 10 to 25 and tasks ranging from 1000 to 5000 task. The experimental results and analyzing the performance of the proposed AMCLBT will be presented in the remainder of this section.

Figure 6 shows the average number of fog nodes are used to perform different number of tasks by DRAM, QLBA, and AMCLBT. As shown in figure 6, AMCLBT technique uses a number of fog nodes less than DRAM, QLBA. This result indicates that the proposed AMCLBT can consume less resources and energy which is a great benefit for this limited resources environment.

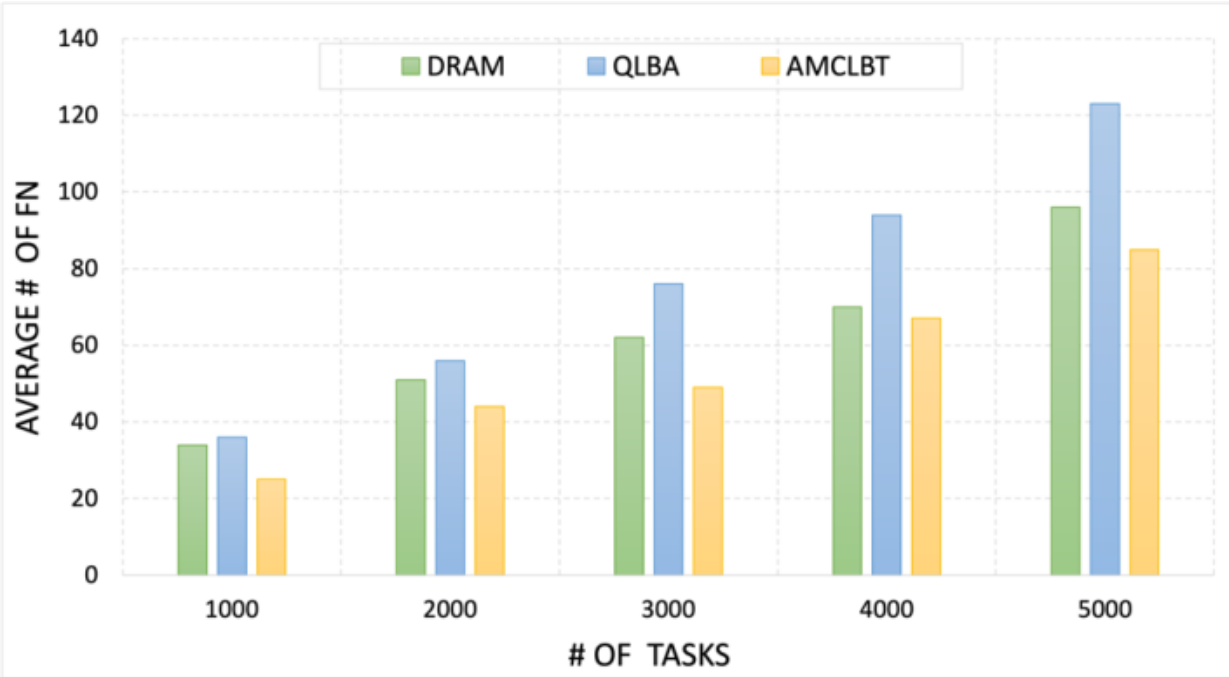

Figure6. Comparison of average number of fog nodes for different number of tasks by DRAM, QLBA, and AMCLBT

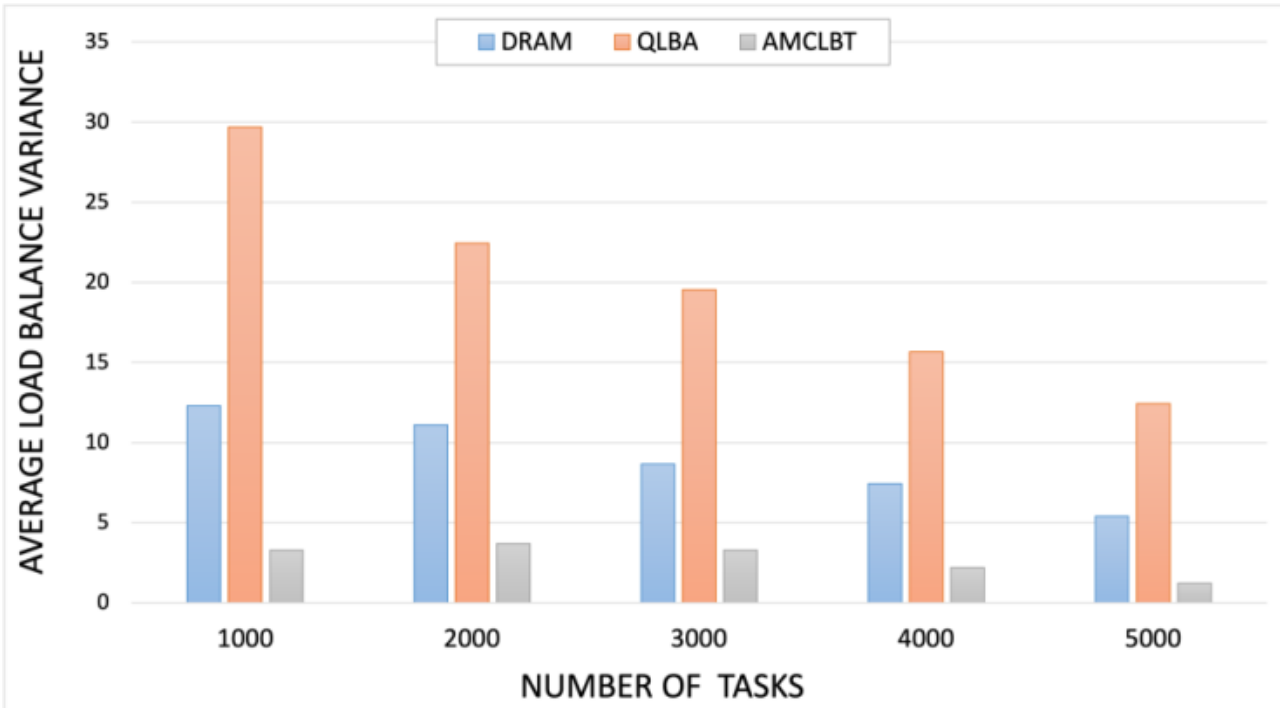

Figure7. Comparison of average load balancing variance for different number of tasks by DRAM, QLBA, and AMCLBT

The proposed AMCLBT tends to minimize the load-balance variance, figure 7 shows the comparison of average load-balancing variance for different numbers of tasks (1000, 2000, 3000, 4000, 5000) for the load-balancing techniques DRAM, QLBA, and AMCLBT. As shown in figure7, the proposed AMCLBT is superior to the other methods, for example when anumber of tasks is 3000 the average load balancing variance for AMCLBT is near to 3 x $10^{-2}$ , whereas DRAM is near 8.5 x $10^{-2}$.

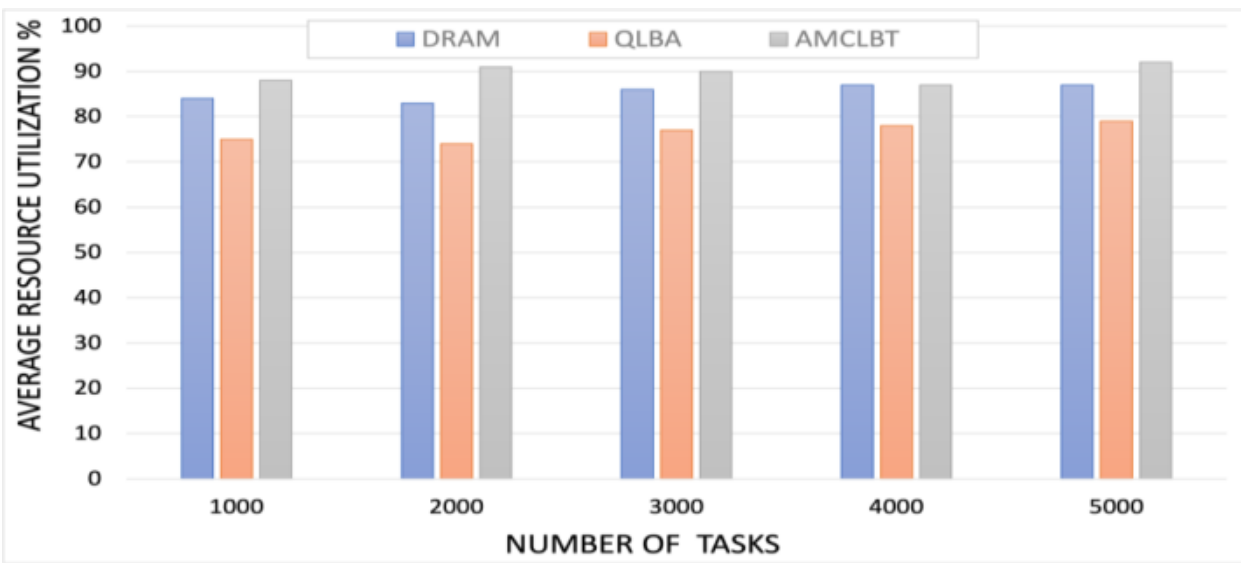

Figure8. Comparison of average resource utilization for different number of tasks by DRAM, QLBA, and AMCLBT

Figure 8 shows the comparison of average resource utilization for different numbers of tasks (1000, 2000, 3000, 4000, 5000) by DRAM,QLBA, and AMCLBT. As shown in figure 8 the average resource utilization of AMCLBT is 90% average and DRAM is 83% average, while QLBA is 75% average. So, the proposed AMCLBT achieved the highest resource utilization among the mentioned techniques.

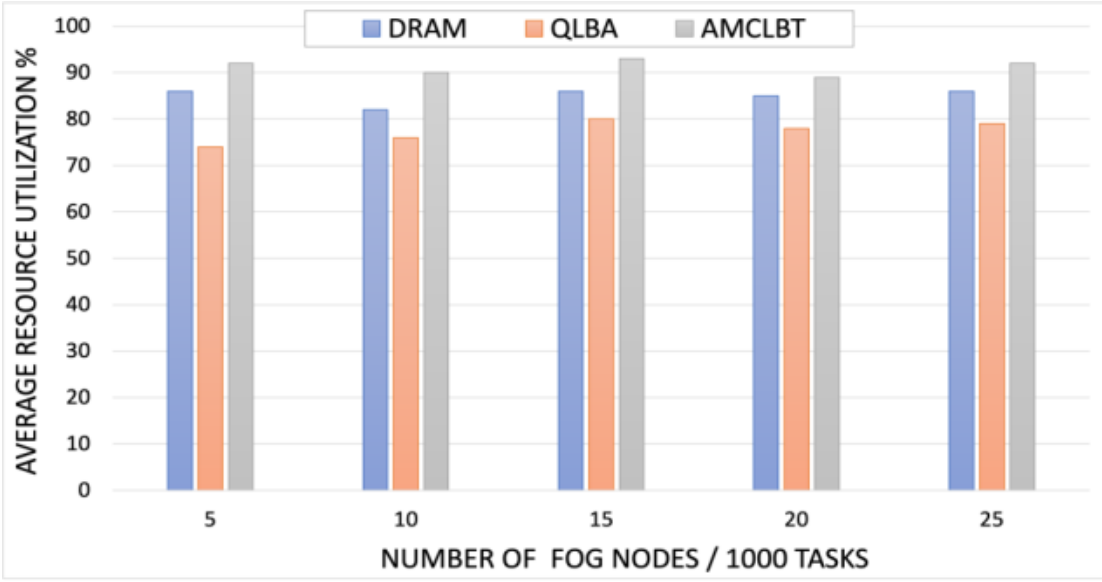

Figure9. Comparison of average resource utilization for different number of fog nodes executing 1000 tasks by DRAM, QLBA, and AMCLBT

The previous comparison results are conducted for different numbers of fog nodes (5, 10, 15, 20, 25) carrying out 1000 tasks to verify previous resource utilization results. The results of this comparison are shown in figure 9, where the proposed AMCLBT technique achieves the highest resource utilization of all the compared techniques.

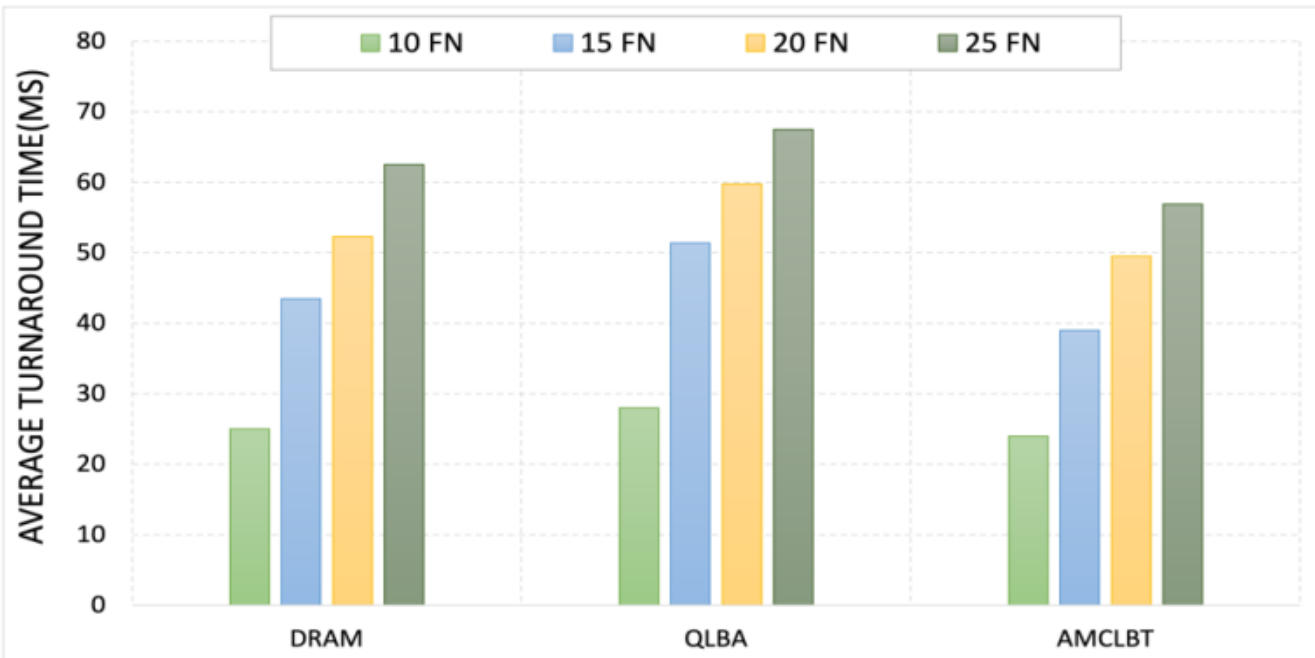

Figure10. Comparison of average turnaround time for DRAM, QLBA, and AMCLBT using different number of fog nodes 10, 15, 20, and 25 executing heterogeneous tasks

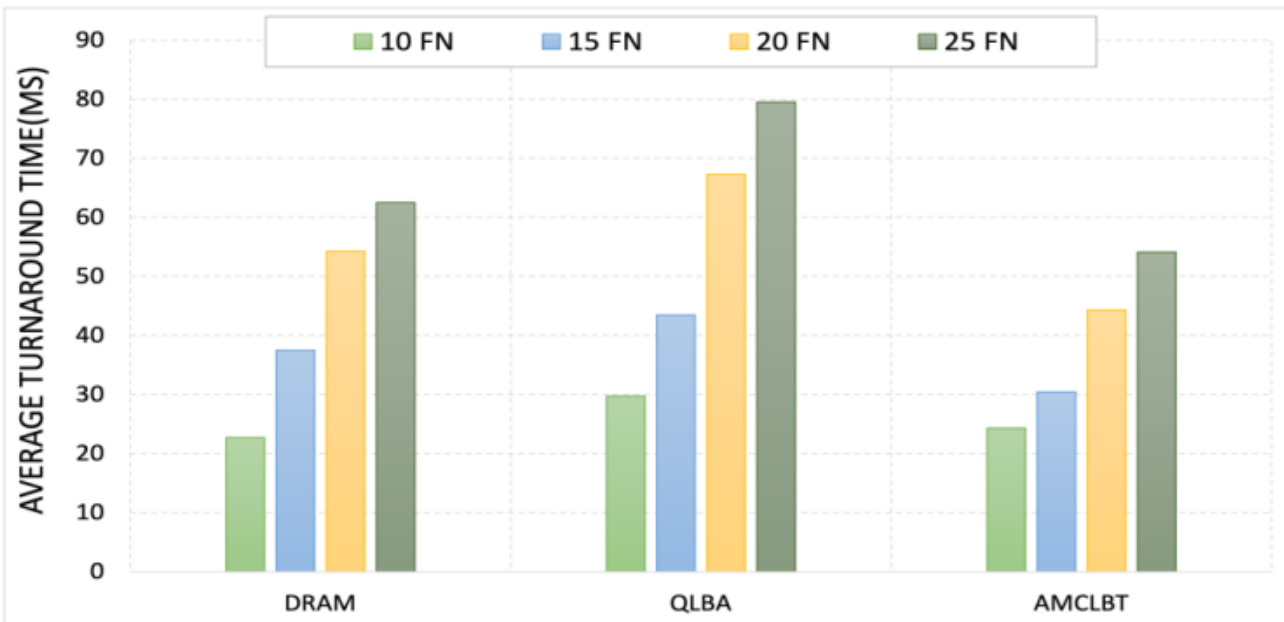

Figure 11.Comparison of average turnaround time for DRAM, QLBA, and AMCLBT using different number of fog nodes 10, 15, 20, and 25 executing homogeneous tasks

Figures 10 and 11 illustrate the average turnaround time for DRAM, QLBA, and AMCLBT with varying numbers of fog nodes (10, 15, 20, and 25) performing heterogeneous and homogeneous tasks, respectively. In a fog computing environment, the turnaround time is defined as the time between task submission and completion. As shown in these figures, the AMCLBT turnaround time for various numbers of fog nodes is less than the other two techniques for both heterogeneous and homogeneous tasks.The performance of the proposed technique is superior to the most recent approaches DRAM and QLBA in several aspects, as demonstrated by the simulation results. The use of fuzzy multi-objective techniques AHP and TOPSIS, which find out the best solution values for several desired objectives, is the factor that leads to this enhanced performance. Using outstanding parameters that affect resource utilization and load balancing between fog devices is a further essential step.

## 9. CONCLUSION

This paper investigates the issue of load balancing in an IoTs-fog environment and its causes and proposed an adaptive multi-criteria-based load balancing called AMCLBT. The proposes AMCLBT, is a multi-objective technique that attempts to address the problem while considering multiple objectives such as available processing, available cache memory, available memory, available bandwidth, available storage, and available energy. These objectives determine the selection of the optimal fog device to conduct IoT tasks while ensuring a fair distribution of tasks on fog devices and maximizing resource utilization. So, these objectives should be considered when solving these types of problems. In addition, AMCLBT uses hybrid multi-criteria decision-making methods such as fuzzy AHP and fuzzy TOPSIS to improve load balancing in an IoTs-fog situation. The simulation results have proven the efficiency of the proposed AMCLBT in solving the problem of fog device load balancing in this environment of limited resources.In the IoTs-Fog environment, resource allocation and load balancing are challenging and active research topics with many related areas. We are investigating into several ways to improve these issues using some evolutionary algorithms. These algorithms might offer an effective method to address these problems.In future work, some of the existing available datasets will be studied to evaluate the proposed algorithm and the impact of data-exchanging procedures will be taken into account.

### CONFLICT OF INTEREST

The authors declare no conflict of interest.

### ACKNOWLEDGMENTS

We would like to thank all authors for their contributions and the success of this manuscript and all editors and anonymous reviewers of this manuscript.